\documentclass[graybox]{svmult}

\usepackage{mathptmx}       
\usepackage{helvet}         
\usepackage{courier}        
\usepackage{type1cm}        
\usepackage{makeidx}         
\usepackage{graphicx}        
\usepackage{multicol}        
\usepackage[bottom]{footmisc}
\usepackage{amsmath}

\makeindex             

\begin{document}

\title*{Stochastic Tools for Network Intrusion Detection}
\author{Lu Yu and Richard R. Brooks}
\institute{Lu Yu \at Clemson University, Clemson, SC, 29634 \email{lyu@g.clemson.edu}
\and Richard R. Brooks \at Clemson University, Clemson, SC, 29634 \email{rrb@g.clemson.edu}}
%
%
\maketitle

\abstract{With the rapid development of Internet and the sharp increase of network crime, network security has become very important and received a lot of attention.  We model security issues as stochastic systems. This allows us to find weaknesses in existing security systems and propose new solutions. Exploring the vulnerabilities of existing security tools can prevent cyber-attacks from taking advantages of the system weaknesses. We propose a hybrid network security scheme including intrusion detection systems (IDSs) and honeypots scattered throughout the network. This combines the advantages of two security technologies. A honeypot is an activity-based network security system, which could be the logical supplement of the passive detection policies used by IDSs. This integration forces us to balance security performance versus cost by scheduling device activities for the proposed system. By formulating the scheduling problem as a decentralized partially observable Markov decision process (DEC-POMDP), decisions are made in a distributed manner at each device without requiring centralized control. The partially observable Markov decision process (POMDP) is a useful choice for controlling stochastic systems. As a combination of two Markov models, POMDPs combine the strength of hidden Markov Model (HMM) (capturing dynamics that depend on unobserved states) and that of Markov decision process (MDP) (taking the decision aspect into
account). Decision making under uncertainty is used in many parts of business and science. We use here for security tools. We adopt a high-quality approximation solution for finite-space POMDPs with the average cost criterion, and their extension to DEC-POMDPs. We show how this tool could be used to design a network security framework.}

\section{Intrusion Detection System}\label{sec:pomdp-ids}
Intrusion detection systems continuously monitor the computer system or network and generate alarms to inform the system administrator of suspicious events.	IDSs are now considered a necessary addition to the security infrastructure of an organization~\cite{Scarfone:2007}.  The objective of intrusion detection is to detect malicious activities, and accurately differentiating them from benign activities.  According to the Common Intrusion Detection Framework (CIDF)~\cite{Tung:1999}, a general IDS architecture has four modules:
\begin{itemize}
\item Event-box (E-box) sensors monitor and collect information about the target system.
\item Database-box (D-box) stores information from the E-box.
\item Analysis-box (A-box) analyzes data stored in D-box and generates alarms if necessary.
\item Response-box (R-box) implements countermeasures to thwart malicious intrusions.
\end{itemize}

The primary classes of detection methodologies include \textit{signature-based detection}, \textit{anomaly-based detection} and \textit{stateful protocol analysis}~\cite{Scarfone:2007}.  IDSs that employ signature-based detection identify attacks by comparing existing signatures of known attacks with the stored network traffic.  When a match is found, IDSs will trigger the corresponding countermeasure to counteract the detected intrusion.  Signature-based detection provides accurate detection results for well-specified attacks and effective known countermeasures can be taken.  The major drawback of signature-based detection is its inability to detect new, unknown attacks.  With new attacks appearing continuously, signature-based detection techniques suffer from high false negative (FN) rates.  The anomaly-based detection has the potential to detect new types of attacks by estimating the deviation of observed information from the predefined baseline of ``normal''.  However, there still exist several significant issues regarding anomaly-based detection, including high FP rates, low throughput but high cost, absence of appropriate metrics, etc~\cite{Teodoro:2009}.  Stateful protocol analysis may provide more accurate detection results than the anomaly-based detection, but is much more resource-intensive due to the complex analysis and overhead generated from state tracking~\cite{Scarfone:2007}.  Typically, the more an IDS's detection accuracy can be improved from the default configuration, the less efficient it is.  As a result, continuous monitoring may cause excessive computation bandwidth, which is undesirable for any computer system and network.

In addition, various intrusion prevention technologies have been implemented in IDSs, such as logging off the unauthorized user, shutting down the system, or reconfiguring the network if possible~\cite{Zheng:2009}, etc.  Despite strengthening the security of the information and communication systems, the intrusion prevention capabilities entail high cost in terms of energy and host resources.

It is not hard to see from the above introduction that, as a popular and effective tool against cyber-attacks by guarding the system's critical information, the resource cost of IDSs must be taken into account.  IDS scheduling is needed to balance between security performance and resource consumption.

There are three primary types of IDS, namely network-based IDS (NIDS), host-based IDS (HIDS) and stack-based IDS (SIDS).  We choose HIDS since HIDS can be selectively deployed on critical machines, such as management servers, data servers and administrator consoles, etc.

\section{Honeypot}
Honeypots are needed to supplement IDSs in the proposed security scheme because they complement most other security technologies by taking a proactive stance.  A honeypot is a closely monitored computing resource used as a trap to ensnare attackers.  As defined by Spitzner in~\cite{Spitzner:2002}, ``A honeypot is a security resource whose value lies in being probed, attacked, or compromised.''  The principal objectives of honeypots are to divert attackers away from the critical resources and study attacker exploits to create signatures for intrusion detection.  The attraction of honeypots to attackers mitigates the threat of malicious attacks and thus helps secure the the valuable information and important services located on the the real targets.

Based on the level of interaction between the honeypots and the attackers, honeypots can generally be divided into the \textit{high-interaction honeypots} and the \textit{low-interaction honeypots}.  Typical examples are honeyd~\cite{Provos:2004} and honeynet~\cite{Spitzner:2002}.  Honeyd allows users to set up multiple virtual honeypots with different characteristics and services on a single machine.  Honeynet monitors a larger and more diverse network when one honeypot may not be sufficient.  It is very complex and expensive to deploy and maintain a high-interaction honeypot because it emulates almost all the activities found in a normal operating system.  Deployment and configuration of a low-interaction honeypot is much easier and cheaper since it only simulates some system services.  ``BitSaucer''~\cite{Adachi:2009} proposed by Adachi et al. is a hybrid honeypot composed of both low-interaction and hight-interaction capabilities.

Honeypots can also be divided into: \textit{research} and \textit{production honeypots}~\cite{Mokube:2007}.  The primary function of a research honeypot is to extract the signatures of emerging attacks, which can be used to improve the detection accuracy of IDSs.  A thorough understanding of the observed traffic data is time-intensive and requires analysts with comprehensive expertise in almost all network-related fields.  Moreover, the deployment of research honeypots provides little benefit in strengthening the system security.  Production honeypots are placed within the production network to mitigate risk.  Most production honeypots are low-interaction honeypots and capture limited information.  Example of production honeypot is Nepenthes~\cite{Baecher:2006}.  Since our scheme is meant to improve security, we use production honeypots in combination with IDSs.

A honeypot mostly does not deal with false positives like IDS since all the services simulated by a honeypot have no production value.  All the traffic that enters and leaves a honeypot is suspicious and should be monitored and analyzed~\cite{Baumann:2003}.  However, not all attempts to access a honeypot are malicious.  For example, a person might mistype the address of a computer and accidentally connect to a deployed honeypot.  As a result, uncertainty is also involved in honeypots scheduling. 

\section{System Model}
Both HIDSs and honeypots detect intrusions and can operate at all times.  However, intrusion detection and system emulation consume a large amount of energy and other resources, including memory, processor usage and disk storage.  We need to balance security performance and the resource cost by scheduling IDS and honeypot activities. The scheduling problem is modelled as a discrete-time process.  In the proposed scheme, more than one HIDS or honeypot can be active during each time period.  

\begin{figure}[!h]
\centering
\includegraphics[scale=0.5]{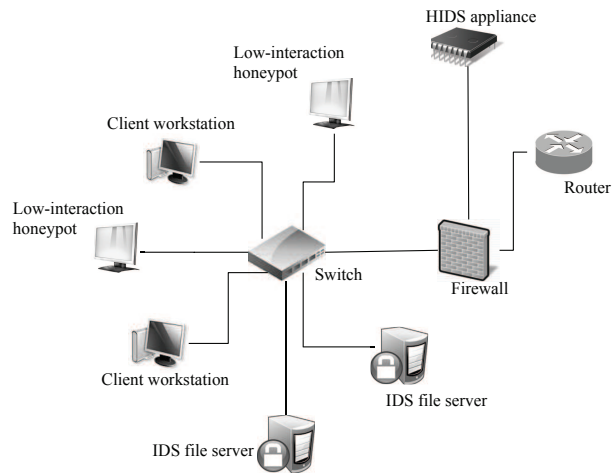}
\caption{Example distributed hybrid security scheme combining HIDSs with honeypots.}
\label{fig:pomdp-ids-honeypot}
\end{figure}

We now consider an example of the hybrid system.  Assume a local area network (LAN) is equipped with $K-H$ HIDSs and $H$ honeypots.  This brings the total number of the security devices to $K$.  Without loss of generality, we also assume the network topology is static and there are more machines than available HIDSs.  Inevitably, only some machines have HIDSs installed.  An example network is shown in Figure~\ref{fig:pomdp-ids-honeypot}.

Suppose each HIDS can operate in three modes: \textit{monitor}, \textit{prevention} and \textit{sleep}, which is the action space for a HIDS.  The HIDS is set to \textit{monitor} for intrusion detection and can sleep for energy saving.  A preventive action might be taken by switching to \textit{prevention} mode if an unauthorized or malicious activity is identified, and consumes more resources.  Similarly, the action space of a deployed honeypot can be specified as \textit{monitor}, \textit{analysis} and \textit{sleep}.  Further analysis will be carried out if traffic anomalies are detected.  Each HIDS makes decision based on local information, which can be modeled as a partially observable Markov decision process (POMDP), which combines the strength of hidden Markov model (HMM)~\cite{yu2013inferring,lu2013normalized} (capturing dynamics that depend on unobserved states) and that of Markov decision process (MDP) (taking the decision aspect into account).

Let $S^{(k)}(t)$ denote the state of an arbitrary device $k$ (HIDS or honeypot) at time $t$, we assume $S^{(k)}(t) = \langle X^{(i)}(t), Y^{(k)}(t)\rangle$, where $X^{(k)}(t)$ represents the security condition and $Y^{(k)}(t)$ represents the resource consumption level.  For instance, the security state can be simply divided into: ``$secure$'' and ``$compromised$''.  If we use the notation $\mathcal{X}$ to denote the state space of $X^{(k)}(t)$, then $\mathcal{X} = \lbrace \textit{secure}, \textit{compromised} \rbrace$.  The state space of $Y^{(k)}(t)$, denoted by $\mathcal{Y}$, includes three consumption levels: $\lbrace low, medium, high \rbrace$.  The correspondence between different consumption levels and operating states are:
\begin{itemize}
\item \textit{low}: The device is not chosen, i.e., in the sleep mode.
\item \textit{medium}: The device is working in the monitor mode.
\item \textit{high}: The HIDS/honeypot is working in prevention/analysis mode.
\end{itemize}

Since the IDSs used are HIDSs, each HIDS only monitors the machine it resides on, ignoring the rest of the network.  As a decentralized control scheme, the decision to activate a certain security device is based on local observations.  To complete the problem, we assume the observation space is identical to the space of security conditions, i.e., $\mathcal{O} = \mathcal{X}= \lbrace ``secure", ``compromised" \rbrace$.  Note that an intrusion alarm does not necessarily mean there is an attack, and vice versa.  Intrusion detection can make two types of errors: false positive (FP) and false negative (FN).  A large volume of FPs result in lots of time wasted on determining whether an alert is an attack when it is actually benign~\cite{Bakar:2005}.  FNs result in security holes due to failing to raise alarms when intrusions occur.   Since the goal of intrusion detection is to precisely differentiate the intrusions from legitimate behaviors, both errors are significant performance indexes of IDSs and have been embodied in the observation probabilities.  For instance, the following observation probability 
\begin{equation}\label{eq:pomdp-dec}
Pr\lbrace O^{(k)}(t+1) = ``compromised" \ | \ A^{(k)}(t) = ``monitor", S^{(k)}(t+1) = \langle secure, medium \rangle \rbrace
\end{equation}
is equal to the FP rate of device $k$.

The local state of each host in the network is closely related to the security posture of the entire network.  The analysis is divided into two cases:  
\begin{itemize}
\item Switching between different modes of an HIDS will change the power-consumption level of the local machine but have no influence on the attack activities in the LAN; 
\item The activation of a honeypot will positively impact the network's operation and security by distracting adversaries away from the valuable resources in the LAN, and accordingly will mitigate the threat posed to the rest of the network.  
\end{itemize}
It is obvious from the preceding analysis that the choice of each agent may affect the state of the entire network.  This makes the decentralized partially observable Markov decision process (DEC-POMDP) a more suitable tool to model the scheduling for the distributed system.  Some might argue that a centralized controller can also be adopted in this case.  However, a common drawback exists in most centralized controls: it may consume a prohibitive amount of bandwidth and instantaneous communication between the agents and the controller.  In addition, possible security breaches are brought in since the transmitted information may be intercepted by the adversaries.  Therefore, the DEC-POMDP is generally more preferable. 

\section{DEC-POMDP Formulation for the Distributed Hybrid Security System Combining IDS $\&$ Honeypot}\label{sec:pomdp-dec-model}
We define the state of the DEC-POMDP formulation at time $t$, denoted by $S(t)$, as the combination of $S^{(k)}(t), i = 1, 2, \cdots, K$. Notationally,
\begin{equation*}
S(t) = [S^{(1)}(t), S^{(2)}(t), \cdots, S^{(K)}(t)] 
\end{equation*}
Similarly, we can write the action of time $t$ as
\begin{equation*}
A(t) = [A^{(1)}(t), A^{(2)}(t), \cdots, A^{(K)}(t)] 
\end{equation*}
We now consider the state transition law.  The state $S(t)$ evolves based on a $(|\mathcal{\mathcal{X}}| \times |\mathcal{\mathcal{Y}}|)$-state Markov decision process.  Let $W$ denote the state transition probability function, then
\begin{equation*}
W(\overline{s}, \overline{a}, \overline{s}') = Pr\lbrace S(t+1) = \overline{s}' | S(t) = \overline{s}, A(t) = \overline{a} \rbrace 
\end{equation*}
where $\overline{s}, \overline{s}' \in \mathcal{S}^{K}$ and $\overline{a} \in \mathcal{A}^{K}$.

The observation probabilities of the DEC-POMDP formulation are slightly different from the standard POMDP.  As indicated in~\eqref{eq:pomdp-dec}, the quantities of the observations probabilities are assigned according to the FP and FN rates of the devices.  Consequently, the observation of each agent only depends on the local information.  It follows that $V^{(k)}(a, s', o')$, the observation probability of device $i$ is given by
\begin{equation*}
V^{(k)}(a, s', o') = Pr\lbrace O^{(k)}(t+1) = o' \ | \ A^{(k)}(t) = a, S^{(k)}(t+1) = s'\rbrace
\end{equation*}
Finally, the immediate reward at time $t$ is defined as the sum of each local immediate reward
\begin{equation}
r(S(t), A(t)) = \sum_{k = 1}^{K}r(S^{(k)}(t), A^{(k)}(t))
\end{equation} 
The values of the immediate rewards are assigned according to the following rules: A successful detection of an attack results in a large reward; On the contrary, an unnecessary further analysis staged due to a misjudgement will cause a large penalty, so will the misdetection of an attack; Furthermore, monitoring is not free and monitoring a secure machine comes at a small penalty.

\section{NLP-based Solution of the DEC-POMDP}

The scheduling model for the hybrid system is a DEC-POMDP.  Thus, we need to augment the POMDP solution method in~\cite{yu2011observable,yu2013applying} to situations of multiple controllers.  As was mentioned in~\cite{yu2012stochastic}, the solution of a DEC-POMDP consists of a set of policy graphs, one for each agent.  Accordingly, the goal is to optimize a set of finte state machines (FSC).  We will show in the followings that the extension of the NLP-based solution in~\cite{yu2011observable,yu2013applying} to DEC-POMDP is very straightforward.  In order to present the algorithm for DED-POMDPs, we make the following assumptions:
\begin{itemize}
\item There are $K$ agents in the DEC-POMDP. 
\item The state space of the DEC-POMDP is denoted by $\mathcal{S}$. Each agent has the same action space $\mathcal{A}$ (``\textit{prevention}'' of an IDS corresponds to ``\textit{analysis}'' of a honeypot) and observation space $\mathcal{O}$.
\item Each agent chooses the actions according to a fixed-size FSC. The set of the nodes in the FSC of agent $k$ is denoted by $\mathcal{N}^{(k)}$.
\item We use the notation ${\overline{n}}$ to denote a vector of length $K$, where ${\overline{n}}(k) \in \mathcal{N}^{(k)}$.  The observation vector ${\overline{o}}$ and the action vector ${\overline{a}}$ are defined likewise.  
\item $x_{k}(n, a)$ and $y_{k}(n, o', n')$ are the control variables of the FSC of agent $k$.
\end{itemize}
The formal representation of the NLP-based solution of DEC-POMDPs satisfying the above assumptions is:
\begin{equation}\label{eq:pomdp-dec-nlp}
\boxed{
\begin{aligned}
& \text{For variables: $\pi_{\overline{n}s\overline{a}}$ and $g^{(k)}(n_k, {o_k}', {n_k}', {a_k}')$,} \\
& \qquad\qquad\qquad \text{where $g^{(k)}(n_k, {o_k}', {n_k}', {a_k}') = x_{k}(n', a')y_{k}(n, o', n')$}  \\
& \qquad \qquad \qquad \text{maximize}\ \sum\limits_{{\overline{n}}}\sum\limits_{s \in \mathcal{S}}\sum\limits_{{\overline{a}} \in \mathcal{A}}\pi_{{\overline{n}}s{\overline{a}}}\cdot r(s,{\overline{a}}^{K}) \\
& \text{Subject to:} \\
& \quad \text{For} \ \forall s' \in \mathcal{S}, \forall \overline{n} \in \bigtriangleup, \forall \overline{a}' \in \mathcal{A}^{K}, \\
& \quad \pi_{\overline{n}'s'\overline{a}'} = \sum_{\overline{o} \in \mathcal{O}^{K}}\sum_{s \in \mathcal{S}}\sum_{\overline{a} \in \mathcal{A}^{K}} \lbrace \pi_{\overline{n}s\overline{a}}\sum_{\overline{o}' \in \mathcal{O}^{K}}P(s, \overline{a}, s')Q(\overline{a}, s', \overline{o}')\prod_{k}g^{(i)}(n_k, {o_k}', {n_k}', {a_k}') \rbrace, \\
& \quad \forall n_k \in \mathcal{N}^{(k)}, \forall {o_k}' \in \mathcal{O}, \sum_{{n_k}'}\sum_{{a_k}' \in \mathcal{A}}g^{(k)}(n_k, {o_k}', {n_k}', {a_k}') = 1, \text{for} \ k = 1, 2, \cdots, K \\
\end{aligned}
}
\end{equation}
The optimal solution to the NLP in~\eqref{eq:pomdp-dec-nlp} provides an optimal set of FSCs of the given size.  The solution representation owes its availability to one critical factor: each agent behaves independently.  That is, all the policy graphs are independent from each other. 

\section{Summary}

This chapter starts with the introduction of the two security technologies adopted in the proposed security scheme.  We choose HIDS, in combination with honeypot with the purpose to integrate the advantages of both tools  in our system.  We formulate the decentralized control of the system as a DEC-POMDP.  In the end of the chapter, we show how to extend the FSC-based POMDP algorithm described in~\cite{yu2011observable,yu2013applying} to solving DEC-POMDP.

\bibliographystyle{spmpsci}
\bibliography{lyuref}

\end{document}